\documentclass{jfm}

\usepackage{graphicx} 
\usepackage[utf8]{inputenc}
\usepackage{color} 
\usepackage{ulem}
\usepackage{amsmath} 
\usepackage{gensymb}
\usepackage{natbib}
\usepackage[lofdepth,lotdepth]{subfig}

\title[Internal Wave Attractors in 3D Geometries]
{Internal Wave Attractors in 3D Geometries : trapping by oblique reflection.}

\author[G. Pillet,  E.V. Ermanyuk, L.R.M. Maas, I.N. Sibgatullin and  T. Dauxois]
  {G. Pillet\textsuperscript{1},  E. V. Ermanyuk\textsuperscript{2,3}, L. R. M. Maas, I. N. Sibgatullin\textsuperscript{5} and  T. Dauxois\textsuperscript{1}}
\affiliation{
  \textsuperscript{1}
 Univ Lyon, ENS de Lyon, Univ Claude Bernard, CNRS, Laboratoire de Physique, F-69342 Lyon, France\\
  [\affilskip]
  \textsuperscript{2}
  Lavrentyev Institute of Hydrodynamics, av. Lavrentyev 15, Novosibirsk 630090,Russia \\
  [\affilskip]
 \textsuperscript{3}
  Novosibirsk State University, Pirogova str. 2, Novosibirsk 630090, Russia \\
 \textsuperscript{4}
  Institute for Marine and Atmospheric Research, Utrecht University, Netherlands\\
 \textsuperscript{5}
 Lomonosov Moscow State University, 119991, Moscow, Russia
}

\pubyear{???} \volume{???} \pagerange{???}
\date{\today}
\begin{document}

\maketitle

\begin{abstract}
We  study experimentally the propagation of internal waves in two different
three-dimen\-sio\-nal (3D) geometries, with a special emphasis on the refractive  focusing
due to the 3D reflection of obliquely incident internal waves on a slope. Both studies are initiated  by ray 
tracing calculations to determine the appropriate experimental parameters.
First, we consider a 3D geometry, the classical  set-up to get simple, 2D parallelogram-shaped attractors  in which waves are forced in a direction perpendicular to a sloping bottom. Here, however, the forcing is of {\textit{reduced}} extent in along-slope, transverse direction.
We show how the refractive  focusing mechanism explains the formation of attractors over the whole width of the  tank,
even away from the forcing region. {Direct numerical simulations confirm the dynamics,
emphasize the role of boundary conditions} {and reveal the
phase shifting in the transversal direction.} 
Second, we consider a long and narrow tank having an inclined bottom,   to simply reproduce a canal.
In this case, the energy is injected in a direction parallel to the slope.  Interestingly, the wave energy ends up forming 2D internal wave attractors in planes  that are transverse to the 
initial propagation direction. 
This focusing mechanism prevents indefinite transmission of most of the internal wave energy along the canal.
\end{abstract}

\section{Introduction}

Internal waves play an important role in  ocean circulation. Generated by tides and winds, they propagate through the oceans and seas, redistributing momentum and energy before dissipating. The mechanism leading to dissipation and mixing remains to be clearly established but at least four possible dissipative processes are regularly invoked and still debated \citep{Kunze2004}: wave-wave interactions and transfer to small scales through triadic resonant instability 
\citep{MacKinnonWinters2005,Alford2007,ARFM2018},  scattering by mesoscale structures \citep{Rainville2006},
or by finite amplitude bathymetry \citep{Peacocketal2009}, reflection on sloping boundaries, especially critical ones \citep{Eriksen1982,DauxoisYoung1999,Nashetal2004}. In this context, the studies of reflections of internal waves on topographies are of particular interest since 
this dissipative mechanism may contribute  to additional  mixing of the ocean. In this paper, 
 we will be particularly interested in three-dimensional reflection processes.

A usual simplification for the propagation of oceanic internal waves is to consider a non-rotating and stably-stratified fluid with a  linear stratification. 
 This is a reasonable first approximation since the ratio of internal gravity wave length over the (internal) Rossby deformation radius  is usually small enough 
at the relevant scales for beam-wise propagating  internal waves and also because  we will consider reflections in closed domains, neglecting propagation
over long distances.
 In this framework, Archimedes' principle is  quantified by the buoyancy frequency $N = \sqrt{({-g}/{\rho_0}){\mbox{d} \overline{\rho}}/{\mbox{d} z}}$, with $g$ the gravitational constant, $\rho_0$ a constant reference density, and $\overline{\rho} (z) $ the unperturbed density field. The dispersion relation of internal waves in such a fluid is given by $\omega = \pm N \sin \theta$, where $\theta$ is the angle of the direction of wave energy  propagation with respect to the horizontal.
 One of the interesting consequences of this  anisotropic dispersion relation is the preservation of the angle of propagation of an internal wave beam upon reflection at a rigid boundary. 

\begin{figure}
\centering
\null\hfill
\subfloat{
\includegraphics[scale=1]{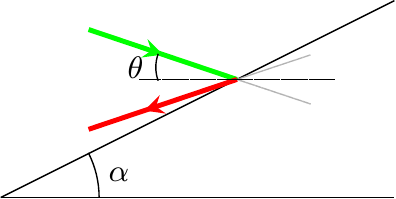}
}
\hfill
\subfloat{
\includegraphics[scale=1]{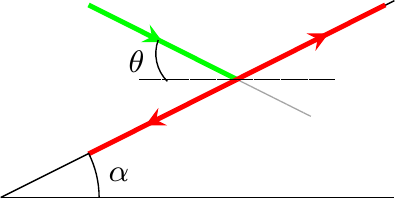}
}
\hfill
\subfloat{
\includegraphics[scale=1]{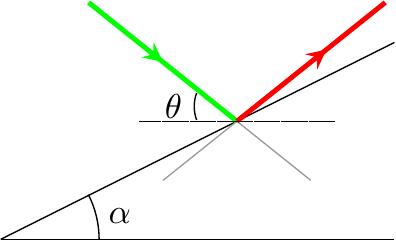}
}
\hfill\null
\caption{ Schematic presentation of the possible  reflections can occur when a {\em linear}  internal wave beam  is propagating in two dimensions. Left: Subcritical reflection. Center : Critical reflection. Right: Supercritical reflection. At the bouncing point, the St Andrew's cross is plotted in grey. The incident and reflected rays are respectively plotted in green and red. }
\label{reflection2D}
\end{figure}

The reflection on planar slopes in two dimensions has been extensively studied since 
\cite{PhillipsBook} and is  now well understood. Depending on the angle of propagation $\theta$ and the angle of the slope$\alpha$,
reflections are divided into three categories. If $\theta < {\alpha}$, the reflection is subcritical, and  the reflected  beam propagates downward (see figure~\ref{reflection2D} left panel). If $\theta > \alpha$, the reflection is supercritical,
the {reflected} beam   propagates upward  (figure~\ref{reflection2D}, right panel).  In the critical  case  $\theta = {\alpha}$, following only the linear predictions, the beam could  therefore be expected to 
propagate both upward and downward (figure~\ref{reflection2D}, middle panel): however, the amplitude of a reflected beam would be infinite.
The singularity of this peculiar critical reflection has been healed by taking into account appropriately
transience, nonlinearity and viscous effects ~\citep{DauxoisYoung1999,Akylas2005}.
 
Given these bouncing linear rules, one can easily compute the path of an  internal wave beam in a given geometry. These trajectories are very different from what we are used to with acoustic
or optical waves, {nearly always} leading to internal wave attractors. In a given geometry, an internal wave attractor is indeed a path toward which all internal waves of a given frequency  will converge: 
this is therefore a limit cycle
although only the \textit{linear} dispersion relation is used. It has been tested through ray tracing that this attracting  structure exists in various geometries~\citep{MaasLam1995,Maas2005,Hazewinkel2011,Brouzetetal2016}. Once the geometry is given, the domain of existence of attractors corresponds to  a large region of the parameter space describing the geometry \citep{MBSL1997,Maas2005,Arnold2017}. These arguments underline the possibility that appropriate
conditions to get wave attractors can
 be encountered  in the  ocean
~\citep{BuhlerMuller2007,Miranda2016}. If this is the case,
 wave attractors  may play an important role in the  
mixing. Indeed, since
the internal waves of a given geometry should converge onto one single path,  the energy on this path when not hindered by viscosity
 will lead  to nonlinear effects and therefore  to mixing events~\citep{BSDBOJ2014,BrouzetEPL2016,BrouzetPRF17}.

Internal wave attractors have been extensively studied theoretically~\citep{MaasLam1995,Ogilvie2005,LamMaas2008}, numerically~\citep{GSP2008,RGV2001} and experimentally~\citep{HTMD2008,HBDM2008} in two-dimensions (2D), but no clear evidence of attractors
 has been found in the ocean yet~{\citep{MandersMaasGerkema2004}}. 
 Internal wave attractors in three-dimensional (3D) geometries have 
been much  less studied \citep{MandersMaas2003,DM2007,Hazewinkel2011}. The aim of this article is precisely 
to study these attractors in 3D geometries. In this paper, we will only focus on the simplest attractors, for which the path formed by the attractor is a
 quadrilateral having  
 one single reflection at the surface and one at the side wall,  a (1,1)-attractor. For the sake of simplicity, in the remainder of the manuscript, we will omit to repeat that we discuss (1,1)-attractors only
 and simply refer to them as `attractors'. 
To simplify further, we will conduct our study through two geometries which slightly vary from the 2D case.
\begin{itemize}
\item {\em The 2D attractor-like set-up}: this geometry is one of the simplest to obtain 2D attractors and has been extensively investigated. It consist{s} of a trapezium with given height $H$, length $L$  and width $W$, and a single sloping wall of  tuneable inclination. Usually, the precise width is  of no influence  and the geometry is considered to be  quasi two-dimensional. In this paper, we  not only take a large value of $W$, to be in a fully 3D case, but also we will apply  the forcing  over a limited stretch in transverse direction only, potentially leading to 3D effects. 
 
\item {\em The canal-like set-up}: a long but narrow  tank is taken to mimic a canal while  a sloping bottom is used to simply {
represent} the most essential feature of the topography. Again here, the slope angle $\alpha$ with the horizontal can be tuned. The waves are forced {by the generator} {in along-canal
} lateral direction, parallel to the slope. {Moreover forcing always takes place over a limited stretch of the tank's width only.}
\end{itemize}

With these two experimental set-ups, we will be able to study carefully the robustness of
internal wave attractors when a third dimension cannot be overlooked.
In section \ref{sec_foc}, we first  discuss the reflection and the refractive  focusing that appear in three dimensions. In section \ref{The2Dattractor-likegeometry}, we present the results of the quasi-2D  set-up, first with simple ray tracing, and then {both} experimentally {as well as numerically}. In section \ref{Thecanal-likegeometry}, we introduce the  experiment with the canal-like geometry where the refractive  focusing leads to counter-intuitive and striking wave trapping in the transverse plane. Likewise, we start with ray tracing predictions before discussing the experimental results.
 Finally, in section \ref{Conclusion}, we conclude and draw some perspectives.

\section{Propagation and focusing in 3D}
\label{sec_foc}

\begin{figure} 
\centering
\null\hfill
\subfloat{
\includegraphics[scale=1]{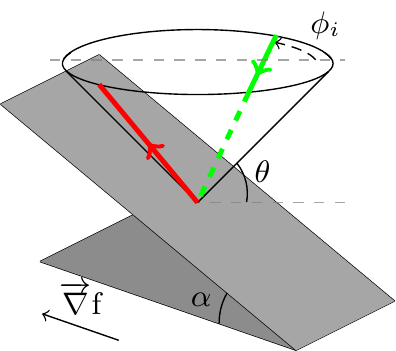}
}
\hfill 
\subfloat{
\includegraphics[scale=1]{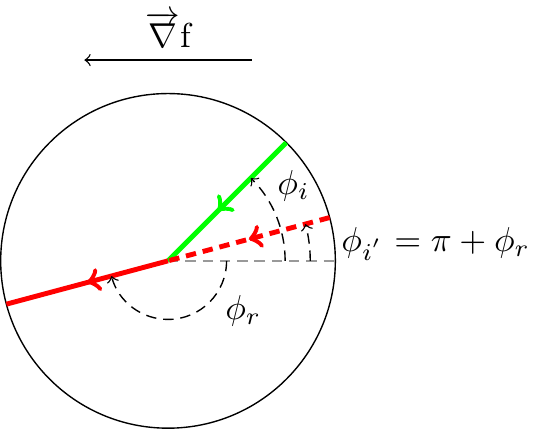}
}
\hfill\null
\caption{Perspective view (left panel)  and top view (right) of the reflection of an internal wave {beam} off an inclined slope {$z= f(x,y)$.} 
{The bottom, inclined at angle $\alpha$ with respect to the horizontal $(x,y)$-plane,} is represented by the inclined grey rectangle, {and its up-slope directed gradient is given by
{the horizontal vector} $\nabla f{= (f_x,f_y,0)}$
}. The internal wave beam propagates along a cone whose inclination, $\theta$, is set by the ratio of wave and stability frequencies.  The incident {(in green)} and reflected {(in red)}  beams make angles $\phi_i$ 
and $\phi_r$ relative to the downslope direction respectively (beams are solid when visible while dotted if not). 
} 
\label{th_ref}
\end{figure}

In 3D geometries, the propagation and reflection laws are more complex than in 2D. The St Andrews' cross, generally used for describing two-dimensional propagation, is transformed into a double cone whose aperture is given by the angle of propagation $\pi/2-\theta$. An internal wave is then constrained to propagate on the cone.  But of course,  besides $\theta$, another angle $\phi$ is needed to describe the position of the ray on the cone (see figure~\ref{th_ref}).

When reflection off a boundary occurs, the reflected ray must stay on the cone, but the horizontal propagation direction of the incident ray, $\phi_i$, measured relative to the downslope direction,  changes. The horizontal propagation direction of   the reflected wave ray, $\phi_r$, can be calculated given that of the incident wave ray, angle $\phi_i$. This occurs  according to the following law of refraction, 
\begin{equation}
\label{law_refl}
\sin \phi_r = \dfrac{(s^2-1)\sin \phi_i}{2s\cos \phi_i + s^2 +1},
\end{equation}
derived for a single reflection  by~\cite{Phillips63}, 
 and reformulated and applied iteratively  by~\cite{Maas2005}.
 {Here, $s(x,y,z) = \tan\alpha/\tan \theta$, represents the local slope of the topography, $\tan \alpha$,
 normalized by the characteristic slope, $\tan\theta$, while the bottom is
 $z =f(x,y)$, whose slope is 
given by $\tan\alpha=[f_x^2+f_y^2]^{1/2}$
and subscripts denote differentiation.}
 This formula is obtained by taking into account the dispersion relation of internal waves  and the impermeability  condition when internal waves reflect from bottom, surface, sidewall or slope. An example of a ``supercritical'' 3D reflection is sketched in  figure~\ref{th_ref}.
 
This law of reflection gives birth to a whole class of behaviour absent in two-dimensional geometries. After bounces on vertical and horizontal walls, the ray may come back on the inclined slope, but this time with a new incident angle $\phi_{i^{'}} = \pi +\phi_r$  measured with respect to the upslope direction. In some cases, this new angle is smaller than 
the incident one, $\phi_i$, meaning that the angle of propagation will converge to $0\degree$ after numerous bounces from surface and sloping bottom. This is the refractive focusing pointed out  by~\cite{Maas2005}. Given the definition of the angle $\phi$, it means that upon multiple surface and subsequent slope reflections  the ray will  eventually propagate perpendicular to  the slope considered.

{Internal wave trapping at a finite distance from the emission point, 
 in a plane transverse to the direction into which these waves are launched, may seem paradoxical.} 
{Upon a focusing reflection, wave length and group speed of a reflecting internal wave decrease. Hence, when the internal wave refracts towards the transverse direction, its energy propagation slows down and, in the ideal fluid description, comes to a halt when that wave length vanishes. This does not imply, however, that the (oscillatory) fluid motion is {confined}, 
in the down-canal  direction; it is merely reoriented upslope. Moreover, when the internal wave scale decreases due to (repeated) geometric focusing, at some point focusing is physically {controlled} by viscous and/or nonlinear processes.}

{The ray advancing into the fluid domain, first propagating completely parallel to the slope, when being diverted upslope, will seemingly slow down. This is because the velocity parallel to the ray will increase in magnitude in cross-slope direction, but not in down-canal direction. Hence, the computation of subsequent reflections will be more and more dominated by these increasing cross-canal speeds until, when the ratio to the down-canal velocity approaches infinity, the ray {comes} to a complete stand-still in down-canal direction.}

\section{The 2D attractor-like geometry}
\label{The2Dattractor-likegeometry}

\subsection{Ray tracing prediction}

In order to have an idea of the path of the internal waves in a 3D geometry, one can simply rely on  ray tracing. 
{In a linearly stratified fluid, internal waves are propagating along straight lines. Given the  coordinates of 
the point from which the wave ray is sent to the slope,  and the initial values for angles ($\phi_0$,
$\theta$),  one can compute the location of the bottom reflection and the local bottom slope, $s$. Then, using Eq. (2.1), 
one can determine angle $\phi_r$ 
of the reflected beam. This allows computation of the location of the next surface reflection, and subsequent 
horizontal propagation direction, $\phi_0'=\phi_r+\pi$.}  
Iterating this process several times, one obtains the path an internal ray beam follows as shown, for example, by~\cite{Rabitti2013,Rabitti2014}. They mainly focused on spherical and spherical  shell geometries but a simple canal in a rotating fluid has also been
studied experimentally and by ray tracing ~\citep{MandersMaas2004}. In each case, they found that, for a given range of parameters, the internal ray, thanks to the focusing effect described in the previous section, ends up forming an attractor which is confined  to a plane and thus identical to 2D attractors. 

\cite{BrouzetEPL2016,Brouzetetal2016} have 
studied 2D attractors in a trapezoidal geometry, which are  easy to set up experimentally. The tank used in these  studies was 350 mm high, 600 mm long and quite narrow,  only 170 mm wide. More importantly, the wave generator was, on purpose, almost as wide as the tank's width. The two dimensionality of the attractors in such a geometry was checked carefully  by~\cite{Brouzetetal2016}. Our first goal is thus to see what happens to this attractor in a tank  that is much wider (W = 800 mm) than the width of the forcing device (150 mm), changing it into a genuine 3D problem.   Indeed, we reused the same generator as for the 2D case. The slope with a tunable angle {$\alpha$}
 is placed along the whole width of the tank (as in the 2D setup).

\begin{figure}
\includegraphics[scale=1]{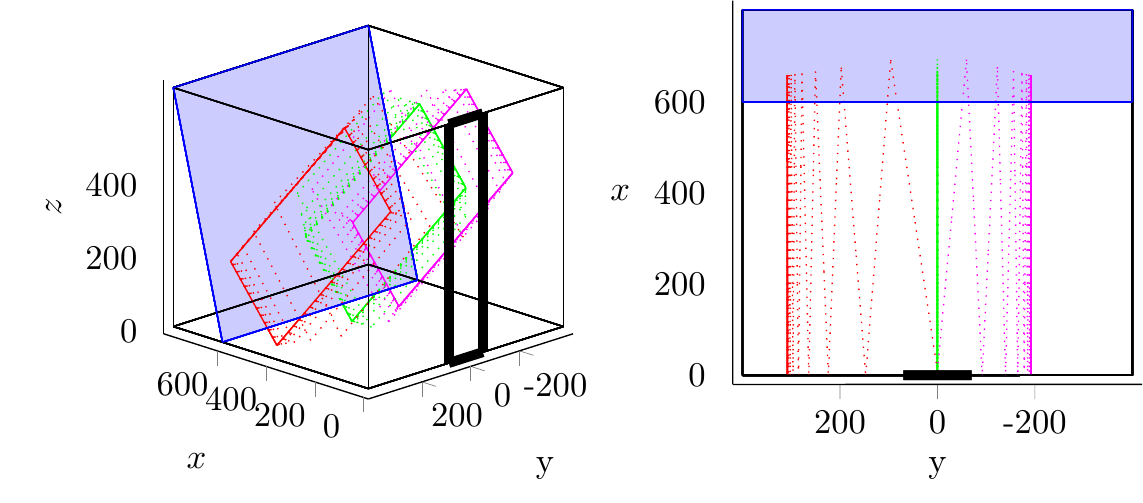}
\caption{The 2D attractor-like geometry: Perspective (left) and top (right) views. The slope is represented by the inclined  blue rectangle, while the generator is sketched by the thick vertical rectangle in the $x=0$ plane{, stretching from $y=-70$ to $y=70$}. In both panels, quantities are in mm. The successive reflections of three different incident rays are then represented with different colors: rays sent with $\phi_0=0\degree$ in green, 
$\phi_0 = 7\degree$ in red and $\phi_0 = -5\degree$ in magenta.  The three rays start at the same initial position, given by $y_0 =W/2$ and $z_0=H/2$. For clarity of the demonstration, we choose the set of input values to ensure that focusing is slow and more visible in the graphic representation. Similar calculations for the experimental set of parameters would yield faster focusing.}
\label{2D_raytracing}
\end{figure}

We performed ray tracing in this geometry. For rays sent exactly along the cross-slope $x$-direction i.e.  $\phi_0 = 0\degree$, we recover exactly the 2D case: as expected, attractors are created in $xz$-planes, with $y$-values identical to   those of the initial rays. This is shown in the left picture of figure~\ref{2D_raytracing}, in which  the green ray represents a wave sent with $\phi_0 = 0\degree$:  the ray path  sticks to   a transverse plane.  

Now if we change the $\phi_0$-value, meaning we send rays with a small angle with respect to the $x$-axis, the red ray path in the left panel of figure~\ref{2D_raytracing} shows  that the trajectory is not 2D anymore. The ray path is clearly 3D, but, due to the focusing effect described in \S~\ref{sec_foc}, the component of the ray along the width is reduced at every bounce against the slope and we end up with a trajectory confined to a  transverse $xz$-plane. The final  $y$-values of these planes depend in a non trivial way on the initial position of the ray beam, but also on  the initial angle $\phi_0$.

The right panel of figure~\ref{2D_raytracing} shows the tank from above so that we can clearly see if the propagation is bounded in a $xz$-plane or not. By contrast, this representation does not allow us to see that the paths in the $xz$-planes are attractors.

 Introducing in the ray tracing program the experimental  parameters, we found that an angular spreading  of only  $\phi_0 = 0\degree \pm 5\degree $ is sufficient
to get attractors created everywhere transversally throughout the tank. Moreover, as the geometry is transversally invariant (only the forcing is not), and the frequency given, the final attractors are all identical,  but in different transverse $xz$-planes. 

In the experimental set-up, even if the waves are forced theoretically with $\phi_0=0\degree$, a diffraction-like  phenomenon can be expected  for the angle $\phi_0$, as obtained previously when this generator was used in a tank significantly wider than the generator~\citep{GostiauxDidelle,Bordes2012}. So we expect to see experimentally attractors along the whole tank width, while the forcing only occurs at its center.

\subsection{ Experiments }
 
\subsubsection{Material and methods}

\begin{figure}
\centering
\begin{minipage}{0.49\linewidth}
\centering
\includegraphics[width=0.8\linewidth]{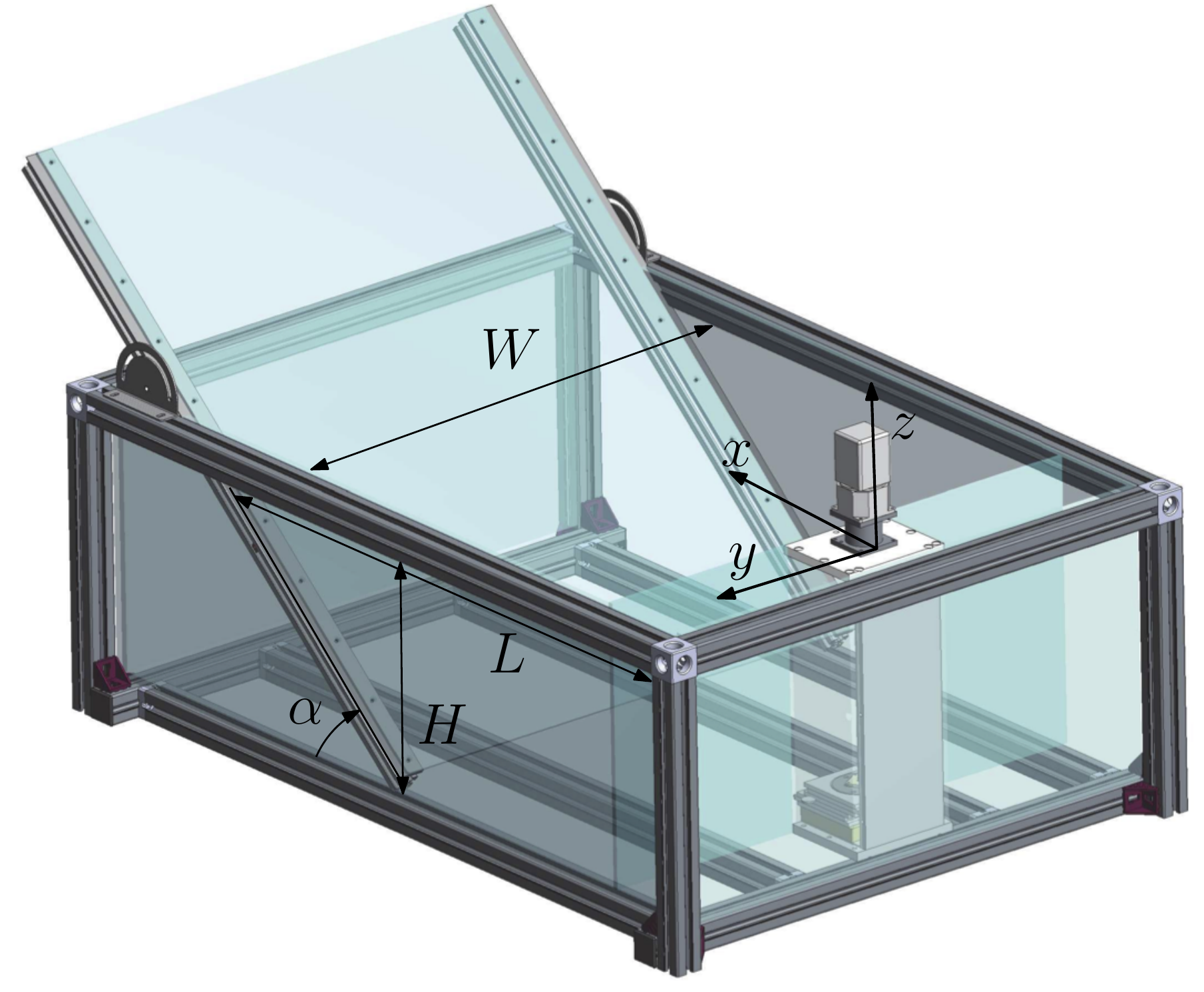}
\end{minipage}
\begin{minipage}{0.49\linewidth}
\centering
\includegraphics[width=0.8\linewidth]{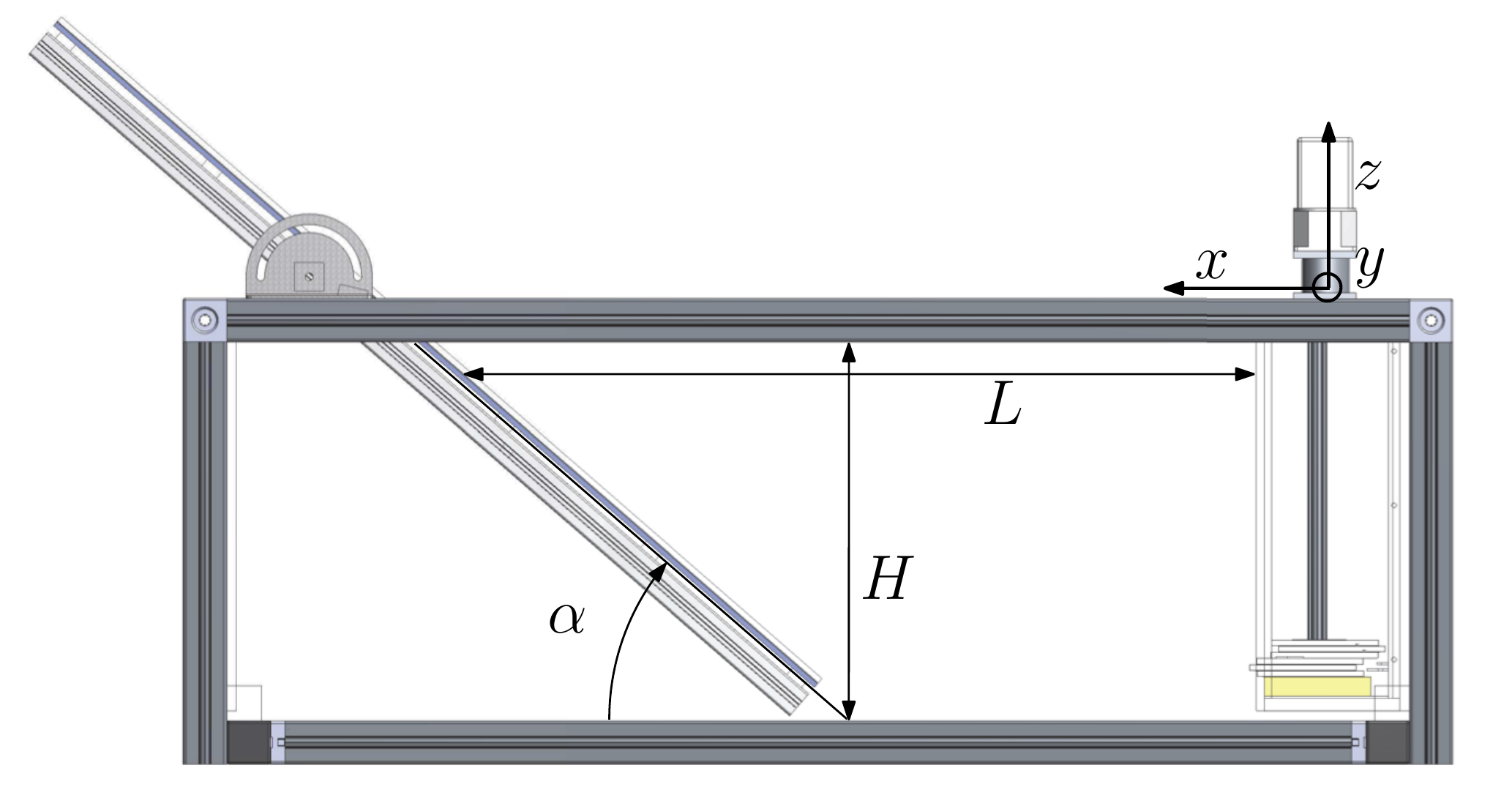}
\end{minipage}
\caption{The 2D attractor-like geometry: Perspective (left) and top (right) views of the experimental set-up. The retractable slope and the wave generator are visible. The parameters used for our experiment are $H = 320$ mm, $L=${570} mm, $W= 800$ mm  {and $\alpha=57\degree$}}
\label{exp1}
\end{figure}

The tank, used for this  first experiment is sketched in figure~\ref{exp1}. Its size is 1200 $\times$ 800 $\times$ 400 mm$^3$. The length,  $L$, at the bottom of the tank in between slope and wave maker   and the slope  inclination $\alpha$ are tuneable. The tank is filled with a  uniformly-stratified fluid using the double-bucket technique~\citep{Oster} with 
salt used as a stratifying agent. The density profile is measured with a conductivity probe attached to a vertical traverse mechanism. The value of the buoyancy frequency is then  evaluated from the measured density profile. {The typical value is $N=0.95 \pm0.05$ rad/s.}

The forcing is created by the  internal wave generator developed by~\cite{GostiauxDidelle},
 studied by~\cite{MMMGPD2010} and later improved by~\cite{Bordes2012}. 
It consists in a stack of 47 plates to which we can change the amplitude and phase of oscillations, that are linked by an Archimedes'  screw. Thus, the time-dependent vertical profile of the generator can be tuned to create different profiles. For this experiment, we force with the first vertical mode of internal waves. The profile is thus given by
\begin{equation}
\zeta(z,t) = a \sin(\omega t)\cos(\pi z/ H) \label{modeone}
\end{equation}
where $a$ is the amplitude of motion of the cams  and $\omega$ is the frequency of the forcing.
                                               
Since we are interested in 3D effects, we cannot use the Synthetic Schlieren method \citep{DalzielHughesSutherland} often used for internal waves measurements when the fluid displacements are two-dimensional to a good approximation. Consequently, here,  internal waves beams are visualised using the PIV technique~\citep{HandbookFM}. We record the displacement of particles lightened by a laser sheet. The particles are silver-coated spheres of size $ 10\,\mu$m and density 1400 kg/m$^3$. The sedimentation velocity is found to be very low compared to the other velocities at play: it will have no consequences. Using a cross-correlation technique~\citep{FinchamDelerce2000}, we can finally  deduce  the velocity field in the tank from these images. The mesh of measurements is found to be sufficient to resolve the small-scale details of the wave field.

In a second treatment, the Hilbert transform is used to reduce noise. This method, first
applied to internal waves by~\cite{MGD2008}, is now widely 
 used  in the internal wave community. The method consists {of} three steps. First, a temporal 
 Fourier transform is performed on the signal. Second, a temporal filter is applied. Finally,
  an inverse Fourier transform makes it possible
  to obtain the filtered field. Generally,  the filter is centered 
  around the forcing frequency $f = {2\pi}/{\omega}$ but it can be used in the case of triadic resonance and/or wave turbulence to study simultaneously several 
   waves of different frequencies ~\citep{BDJO2013,BrouzetEPL2016}. Another common use of the Hilbert transform is to filter  the signal spatially. For instance, one can separate waves propagating toward positive $x$-values from waves propagating toward negative $x$-ones. Both spatial and temporal filtering can be simultaneously applied, as will be done in the remainder of this article.

For this experiment, a laser sheet, created above the tank, illuminates  a  $xz$-plane along the tank. The laser sheet can be easily translated to obtain different slices of the tank, for different $y$-values. 
A computer-controlled video AVT (Allied Vision Technologie) Pike camera with CCD matrix of 1382$\times$1034 pixels is used for video recording. The camera is located at a distance 1600 mm from the tank and operates at a constant frame rate of $2$ Hz, which is sufficient to resolve the significant frequencies of the signal (typically of  the order of 0.1 Hz) and small enough so that the particles stay within the laser sheet between 2 images.

\subsubsection{ Experimental results }

We performed experiments for the 2D attractor-like  geometry with a mode-1 forcing~(\ref{modeone}) with a= 5 mm. For what concerns  the longitudinal  geometry, we have chosen the same parameters as used  by~\cite{BrouzetEPL2016}. However, the forcing applies  only on 150 mm out of 800 mm. 
We expect therefore to see the same attractors in the middle of the tank (where the forcing occurs) and,  according to the  ray tracing prediction, away from the forcing region as well.

\begin{figure}
 \includegraphics[scale=1]{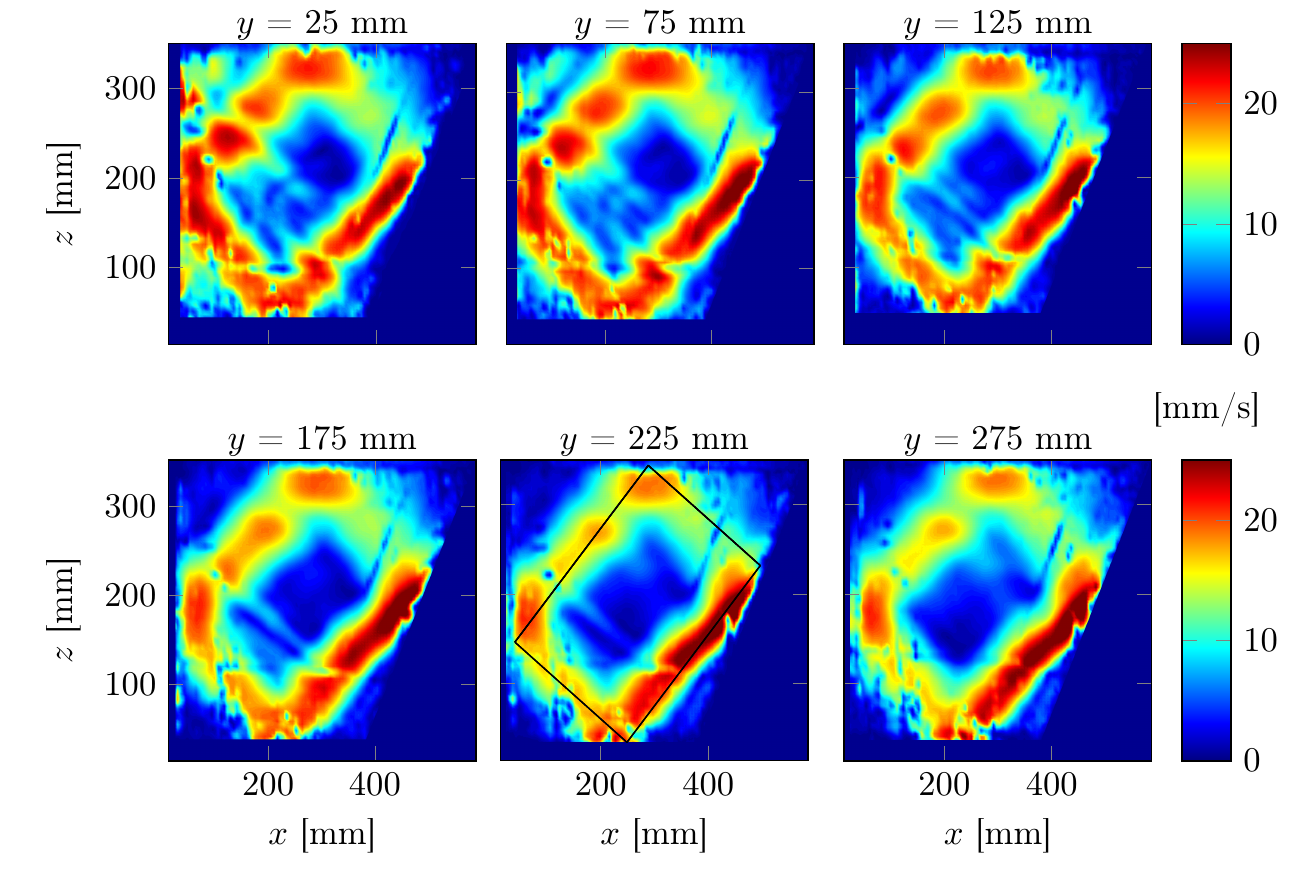}
\caption{Filtered norm velocity for different $y$ positions indicated {above} each panel. As an example, for the bottom middle graph, the ray tracing prediction is plotted with a solid black line, using the {experimental} parameters: $L=360$ mm, $W=800$ mm, $H=320$ mm, $\alpha={57\degree}$, $N = 1.14$ rad.s$^{-1}$ and $\omega = 0.515$ rad.s$^{-1}$.
}
\label{2D_tranches}
\end{figure}

 Let us consider first  the steady state. Once the  generator is started, one  waits  long enough {(typically 100~periods)} to reach  the steady state of the attractor. We then record the displacements of the particles in different $xz$-slices, from the center of the tank to one edge ($y$ varying from 0 to 400 mm). Figure~\ref{2D_tranches} shows  the velocity field obtained after Hilbert filtering at the forcing frequency~$\omega$. 
 
One can see in figure~\ref{2D_tranches} that even if the forcing is only
 centred on a small part of the tank, the velocity field is almost the same 
 everywhere throughout the tank, showing a nice attractor velocity field.
We checked the presence of attractors on both sides of the tank. Because of light absorption, the signal is better when the slice is close to the camera, this is why 
we only show here the slices for positive values of $y$.

Figure~\ref{2D_tranches} shows the existence of  attractors everywhere in the tank, but does not exhibit the appropriate mechanism for the transversal  spreading. One could first argue that attractors away from the forcing can  only be due to  viscous diffusion  from the centre of the tank towards the sidewalls. If so, we would  see an important  time dependence of the appearance of  attractors with   distance from the center of the tank. An order of magnitude estimate of viscous spreading of  the attractor is given by  $y = \sqrt{\nu t}$. For the viscosity of water $\nu \simeq 10^{-6}$ m$^2$s$^{-1}$ and a distance from the center $y=300$ mm, one obtains 
an estimate of the propagation time of the central velocity field of $T \simeq 10^5 $s, or, approximately $T = 10^4 \, T_0$.

\begin{figure}
\centering
\includegraphics[scale=1]{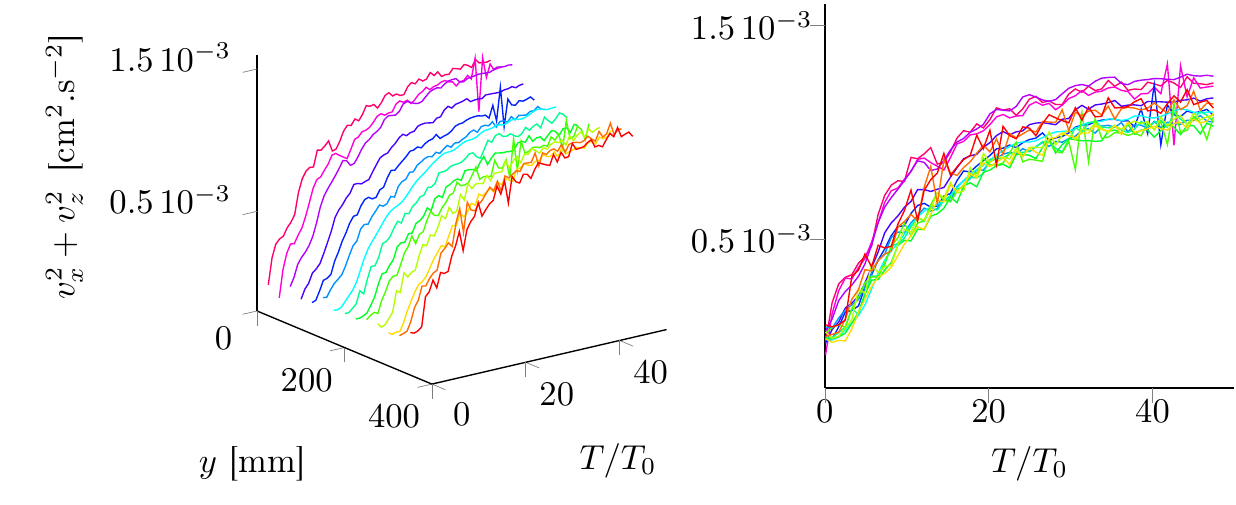}
\caption{Time evolution of the {squared} longitudinal velocity amplitude for different transversal slices of the tank. A different color is used for the 14 different $y$-slices varying from 0 to 350mm. Left: Perspective view. Right: Side view. The oscillations at the forcing frequency have been removed using Empirical Mode Decomposition.}
\label{tranches_color_2}
\end{figure}

We hence performed  a second series  of measurements, focusing on  the transitory regime of the attractors in each slice to see if such a large  shift is observed between the time for  reaching the attractor steady state. For each slice, we started the camera acquisition and the generator at the same time. Before analyzing, we perform an Empirical Mode Decomposition (EMD) on the signal. This method described in \cite{Huang1998} and \cite{Flandrin2006} allows one to decompose the signal in the sum of all the important frequency contributions. In the present experiment, since we are only interested in the transitory regime, it allows us to get rid of high frequencies. 
In this experiment, strong oscillations at the forcing frequency are indeed 
present. As explained
in~\cite{Arnold2017}, these oscillations are due to the propagating or standing nature of the waves and are not of interest here, since they do not modify the timescale of the transitory regime. We then looked at the sum of the squared longitudinal velocity amplitudes, $v_x^2+v_z^2$. Figure~\ref{tranches_color_2} presents its time series for every slice. 
We can clearly see that the steady state is reached after  approximately the same transient 
time for each slice, hence invalidating the intuitive mechanism of attractor spreading along the width of the tank by viscous diffusion. The rapid propagation of  wave energy through the tank and its relative homogeneity in the steady state both strongly support spreading by wave propagation followed by refractive trapping.

\subsection{{ Numerical simulations}} 

\begin{figure}
\centering
\includegraphics[width=0.8\textwidth]{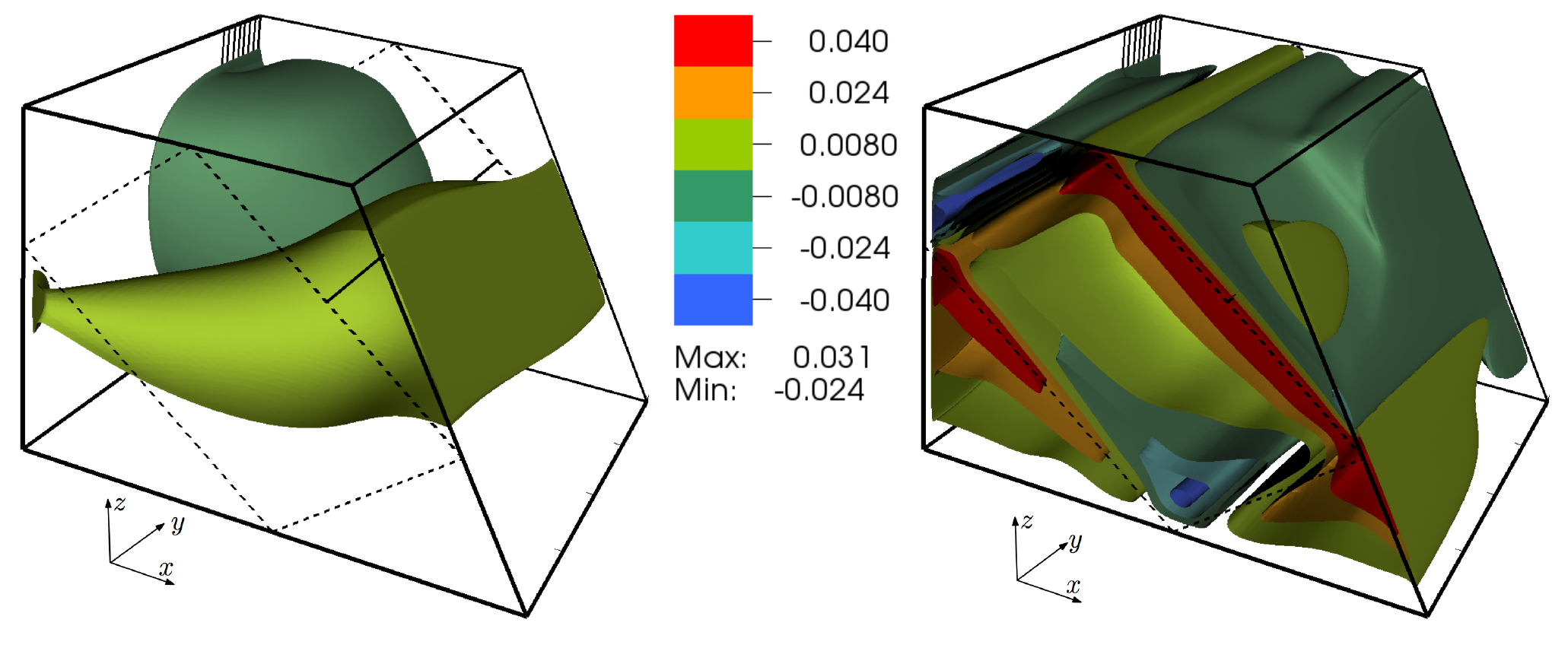}
\caption{
{Perspective views of the attractors with the contours of the vertical velocity at $t=16$~s~(left) and $t=2504.5$~s~(right). The simulation results correspond 
to the following parameters:  $L = 456$ mm, $W = 800$ mm, $H = 308$ mm, $\alpha =60^\circ$, $N = 0.948$~rad.s$^{-1}$, and $\omega=0.589$~rad.s$^{-1}$.
The generator, sketched by the vertical lines at the back of the figure, had an amplitude $a=$0.6~mm.}
{The dashed lines drawn on the sides of the domain correspond to the theoretical skeleton of the wave attractor predicted by the ray tracing. The horizontal solid line connecting the middles of the first rays on the opposite lateral sides shows the location of the transverse cut for which we show below the transverse distribution of the cross-correlations and amplitudes of the wave motion.}}
  \label{fig:X2}
\end{figure}

{To have a complementary view of the above results, we have performed direct numerical simulations. 
We used a well-tested spectral element approach, a version of the Nek5000 code~\citep{Fischer1997,Fischeretal}. Using this code, three dimensional numerical modelling of a quasi two-dimensional laboratory setup, described in~\cite{Brouzetetal2016}, showed excellent correspondence to the experiments (see also~\cite{BrouzetEPL2016}).
Results of direct numerical simulations of the present problem are found to be similar with the experimental results. The initial stage of the transient process of the wave field formation is depicted in Fig.~\ref{fig:X2}(a) with the help of contours of the vertical component of velocity. The amplitude of the wavemaker was chosen to be sufficiently small to ensure the attractor is in the linear regime. Only half of the domain is shown in spanwise direction and, to visualize the internal flow structure, we have cut off the near-wall region $y>390$~mm. The ratio of the wavemaker width to the whole width of the tank is the same as in the experiments. As could be expected, the wave perturbations propagate from the wave maker over a wide range of azimuthal directions. The dashed lines drawn on the sides of the domain correspond to the theoretical skeleton of the wave attractor predicted by the ray tracing. The horizontal solid line connecting the middles of the first rays on the opposite lateral sides shows the location of the transverse 
cut for which we show below the transverse distribution of the cross-correlations and amplitudes of the wave motion. Below, this line is referred to as the ``probe line''. Figure~\ref{fig:X2}{(b)}  presents 
the isosurfaces of the vertical component of velocity after 235 periods of the wavemaker oscillations. Note that we use the same colour bar in both panels of Fig.~\ref{fig:X2}. It is clearly seen, that the isosurfaces form a structure, which is close to a two dimensional one. At the same time, noticeable deviations from a purely 2D structure are visible: interestingly, the velocity may increase in transversal direction towards the lateral side wall.}

\begin{figure}
  \centering
   \includegraphics[scale=0.75]{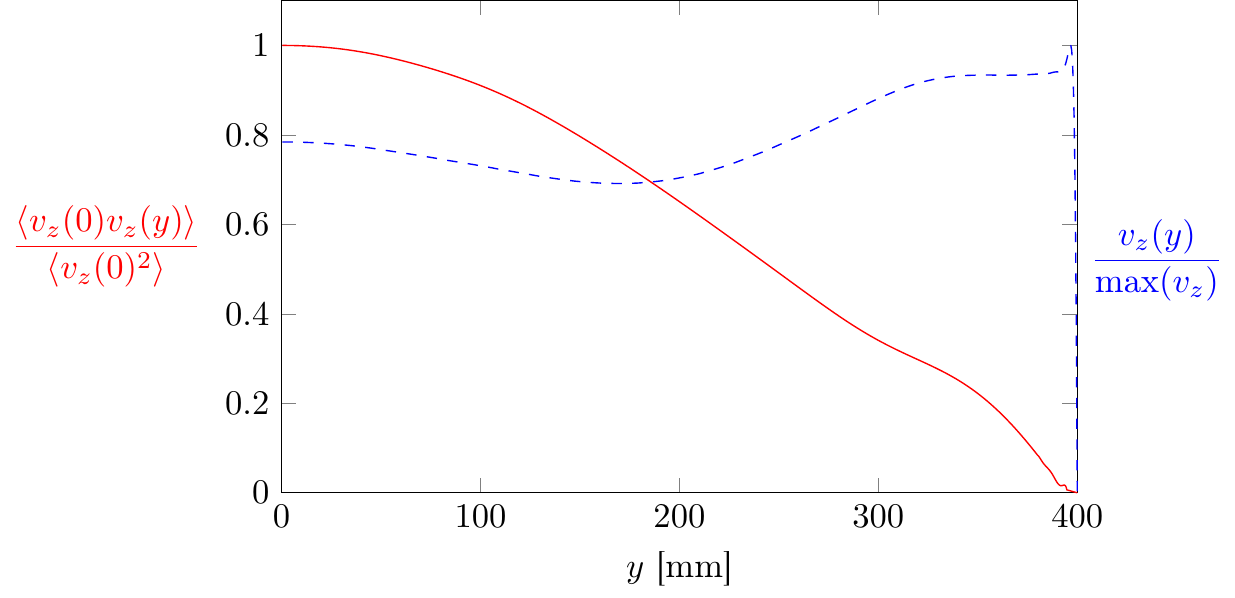}
  \caption{{Correlation 
  ${<v_z(0, {t})v_z(y, {t})>/<v_z(0, {t})^2>}$ (solid line) and normalized velocity 
  $v_z(y)/\max{(v_z)}$  (dashed line) as a function of the transverse coordinate~$y$ for the same parameters as in Fig.~\ref{fig:X2}.}}
  \label{fig:X3}
\end{figure}

{Figure~\ref{fig:X3} shows the correlation 
${<v_z(0, {t})v_z(y, {t})>/<v_z(0, {t})^2>}$ between the vertical velocity at the beginning of the probe line (located at the central vertical plane $y=0$, coming through the middle of the wave maker) and the vertical velocity at a different transversal location $y$ at the  same line. {Time averaging has been performed 
over 102 periods of the  forcing oscillation.} The value of the correlation decreases with the transverse coordinate. The dashed line shows the transverse distribution of the velocity amplitude 
{$v_z(y)/\max{(v_z)}$} along the same line, normalized by the maximum velocity at the probe line. These pictures as well as the analysis of three-dimensional isosurfaces reveal that the velocity amplitude has a non-monotonous dependence on the transverse coordinate. It reaches the maximum close to the lateral wall before falling down to zero due to the no-slip boundary condition. The behavior of the correlation coupled with the behavior of the velocity amplitude allows us to conclude that there is also a phase shift between the wave motions at different transverse locations.}

\begin{figure}
  \centering
  \includegraphics[scale=0.75]{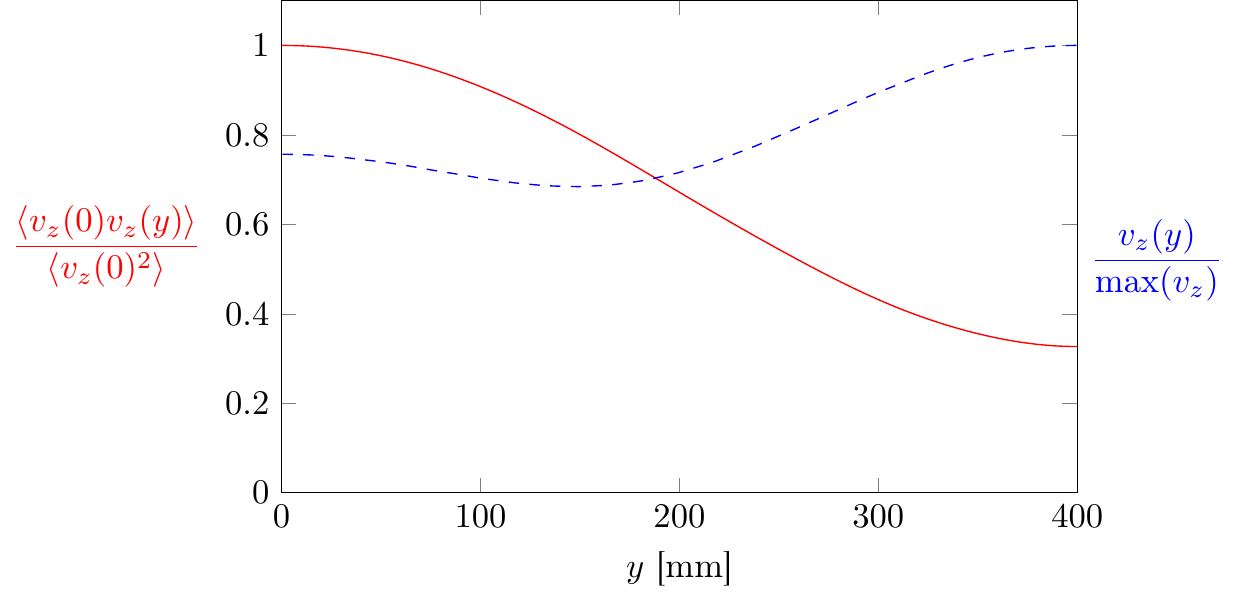}
  \caption{{Same as Fig.~\ref{fig:X3} with stress free, instead of no-slip boundary conditions {at lateral walls}.}}
  \label{fig:X4}
\end{figure}

{To isolate the effect of the wall friction on the lateral wall, we present in Fig.~\ref{fig:X4} the results of similar calculations performed for a stress-free boundary condition at the lateral wall. By comparing the results of simulations with the no-slip (Figs.~\ref{fig:X2} and Fig.~\ref{fig:X3}) and stress-free boundary conditions (Fig.~\ref{fig:X4}) at the lateral wall, we note that the latter case exhibits a more regular and smooth dependence of the calculated quantities on the transverse coordinate: in particular,  there is no localized sharp maximum of the velocity amplitude close to the {free-slip} lateral wall. The shapes of the sharp maxima observed in calculations with the no-slip condition at the lateral wall (see Fig.~\ref{fig:X3}) are reminiscent of the “cat ear” velocity profiles calculated in \cite{Brouzetetal2016} where such features were attributed to the effect of localized recirculating flows. Applying standard boundary layer theory as in \cite{Beckebanze2018} should
 provide an elaborate description of the viscous dissipation in the interior shear layer, as well as at the rigid boundaries. Modification of the geometry and viscosity in calculations produce a variety of different amplitude and phase distributions, which will be described in a separate paper.}

\section{The canal-like geometry}
\label{Thecanal-likegeometry}

\subsection{Ray tracing prediction }
 
The second experimental set-up is of particular interest since it echoes some in situ measurements. While internal waves have been measured close to the mouth of the Laurentian Channel   they appear to be of very low intensity  \citep{SaintLuarentAttenuation}. This is surprising as internal wave dissipation in the ocean is known to be small, normally allowing
internal waves to propagate thousands of kilometers before vanishing \citep{Alford2003}. Here this occurs even though internal tides are known to be generated at the land-locked head of the Channel \citep{Cyr2015}.
A possible explanation for
 this phenomenon 
could precisely be the refractive focusing mechanism investigated here, especially if tidal energy could now form attractors, as in figure~\ref{exp1}, that are trapped on \textit{transverse} planes.
This is why we will study a model geometry of this physical situation, looking for the potential presence of these attractors.

\begin{figure}
\centering
\includegraphics[scale=1]{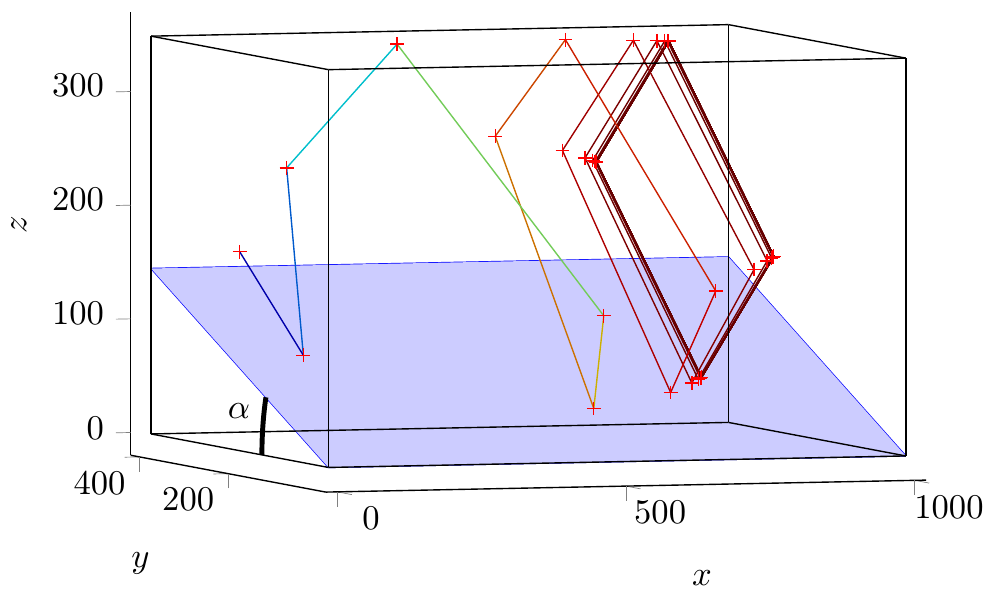}
\caption{Trajectory of a single internal wave beam propagating in the canal-like geometry
filled with a linearly stratified fluid. The beam is sent {downwards} {(in the negative $z$-direction)}  from the plane $x=0$ with  $\phi_0 = 90\degree$ 
(i.e. into the positive,  {along-tank} $x$-direction). Quantities are in mm. 
}
\label{foc_lente}
\end{figure}

Figure~\ref{foc_lente} presents such a set-up for which an internal ray beam path leads to 
an attractor
in a plane transverse  to the  along-slope, down-canal direction into which rays are initially launched. A sloping bottom  is inserted in a parallelepiped tank. It makes an angle $\alpha$ with the horizontal. Starting from an initial point on the wall $x=0$, an internal wave beam of a given frequency~$\omega$ launched with an angle $\phi_0 = 90\degree$ will reflect several times on the different walls as visible in figure~\ref{foc_lente}. 
At each bounce off the inclined slope, the value of $\phi$ changes and converges to~$0\degree$, $i.e.$ onto a  $yz$-plane, transverse to the initial direction, which we call the focusing plane. In the remainder of this paper, we will index these focusing planes by their asymptotic $x$-value, which we call $x_{\infty}$.
Moreover, as the transversal geometry of the canal has one inclined slope, it can lead to an attractor~\citep{Maas2005}.
One finds out that, for a large range of angles $\alpha$, frequency ratios $\omega/N$,  and ray launching positions, attractors can be created. 
In nearly all these cases, the ray will eventually not propagate downstream any longer (for exceptional  'whispering gallery' type  waves, that escape trapping, see~\cite{Maas2005,DM2007}). Consequently, one can identify a zone of propagation of the internal wave ray, followed by a zone of trapping, and  finally a zone without any internal waves.

\begin{figure}
\centering
\includegraphics[scale=1]{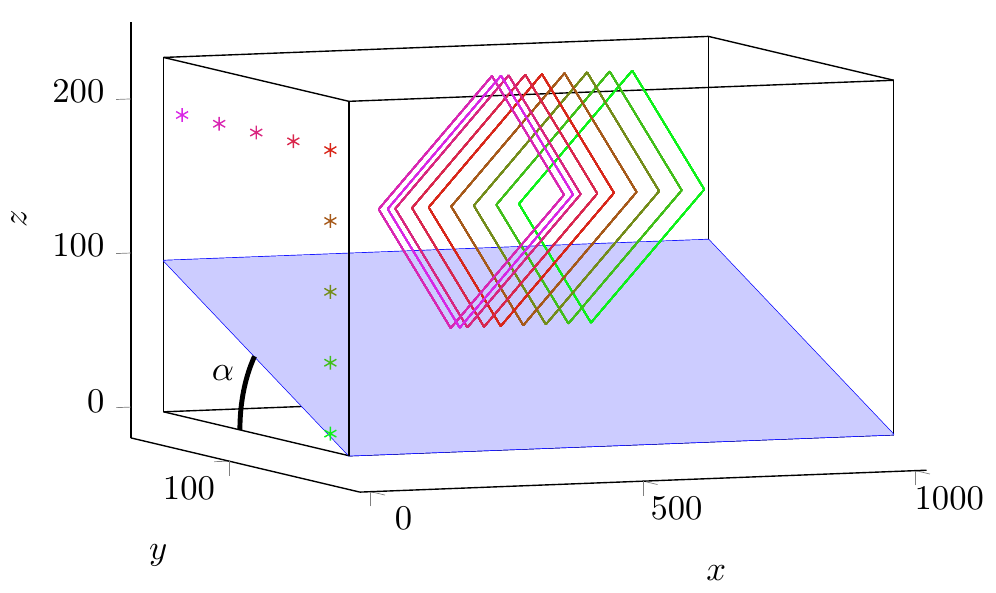}
\caption{Final steady paths launched from different positions in the $x=0$ plane, represented by the different $\ast$-symbols. The rays are launched downward with $\phi_0=90\degree$. The color is the same
 for  the initial point and the corresponding steady path.  Quantities are in mm.
 }
\label{pls_attract}
\end{figure}

A careful analysis shows that for given values of $\alpha$ and $\theta$, the focusing plane depends on the initial angle $\phi_0$ with which the ray is launched and on its initial position $(x_0,y_0, z_0)$. To give an idea of the initial position dependency of  $x_\infty$,  figure~\ref{pls_attract} presents the different steady paths created from several initial positions, all located in the same $x=0$ transverse plane.  The initial positions are represented by the $\ast$-symbols while the steady paths correspond to the coloured lines. One realises that every point leads to a different attractor, each of them lying in a different transverse plane. 

\begin{figure}
\centering
\includegraphics[scale=1]{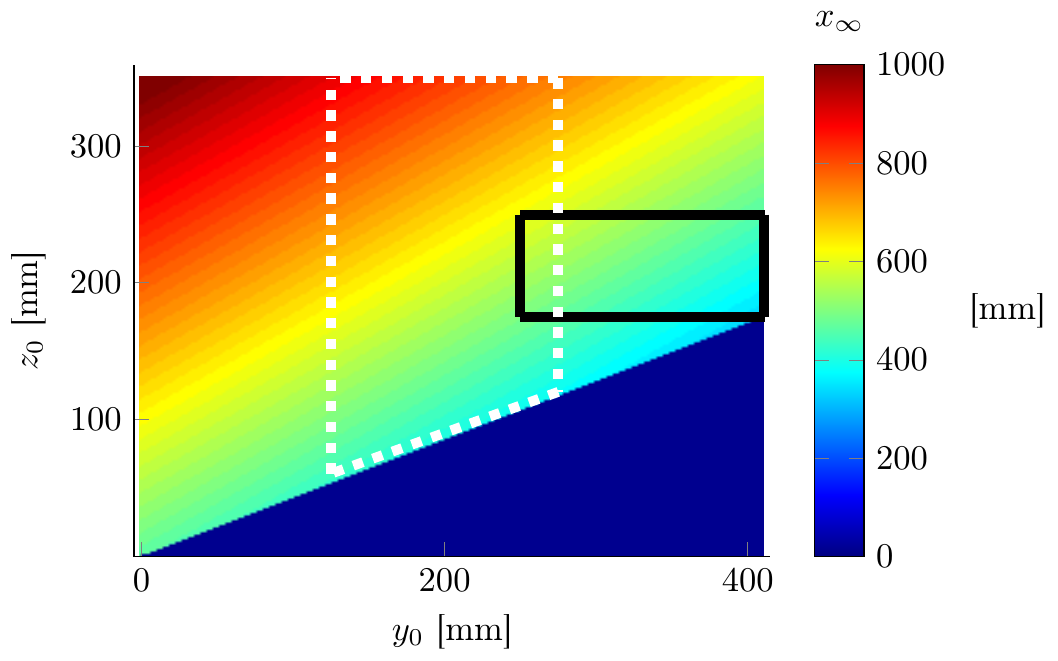}
\caption{For every initial point $(y_0,z_0)$, the $x$-position of the focusing plane is plotted. The parameters for this computation are $\alpha=23\degree$, $\theta=32\degree$, $H=360$ mm, $L=1000$ mm and $W=410$ mm. The rays are launched downward with $\phi_0=90\degree$. The solid black (respectively dotted white) contour represents 
the region  where the plane-wave (respectively mode~1) profile is forced. The
blue triangle corresponds to the uninteresting region below the slope.}
\label{map_foc}
\end{figure}

In order to have an idea of where to expect the attractors, one can compute, for a given 
set of parameters ($\alpha$, $\theta$, $\phi$),
the position map of the focusing plane as a function of $y_0$ and $z_0$. This map  presented in figure~\ref{map_foc} is computed for a set of parameters used in the experiment  described below.
It emphasizes that depending on the region of forcing, attractors can be created close to or far away from the generator. Please note already that in the experiments,  forcing of the fluid 
will be applied over an initial bounded $(y_0,z_0)$-region. In doing so, one can thus expect to find attractors only in a given region of space. 

\subsection{Experiments}

The aim is to reproduce this refractive  focusing predicted by  ray tracing,
using a tank of 1200 $\times$ 410 $\times$ 400 mm$^3$. 
The tank is filled up to 360 mm above the flat bottom and, due to
the generator's thickness, the slope is only 1000 mm long. We chose the origin, $x=0$, 
 at the edge of the wave maker. 
Indeed, when experimentally studying narrow attractors, branches are not often easily distinguishable spatially.
Another constraint is having a propagation angle sufficiently different from the slope angle, so that the first attractor branch will not be affected by viscous effects along the slope. These experimental difficulties are restricting a lot the parameter space, leaving us few possibilities to obtain reliable data. 
The present set-up takes care of this issues by considering
 a tank that is  less wide than the previous one, shown in figure \ref{exp1}. The transverse geometry is taken square shaped, except for an additional slope of angle $\alpha= 23\degree$  that is
put at the bottom of the tank. 
Thus  one may expect to create approximately square-shaped attractors 
that are
easier to observe and to analyse. 
 
\begin{figure}
\centering
\begin{minipage}{0.49\linewidth}
\centering
\includegraphics[width=0.9\linewidth]{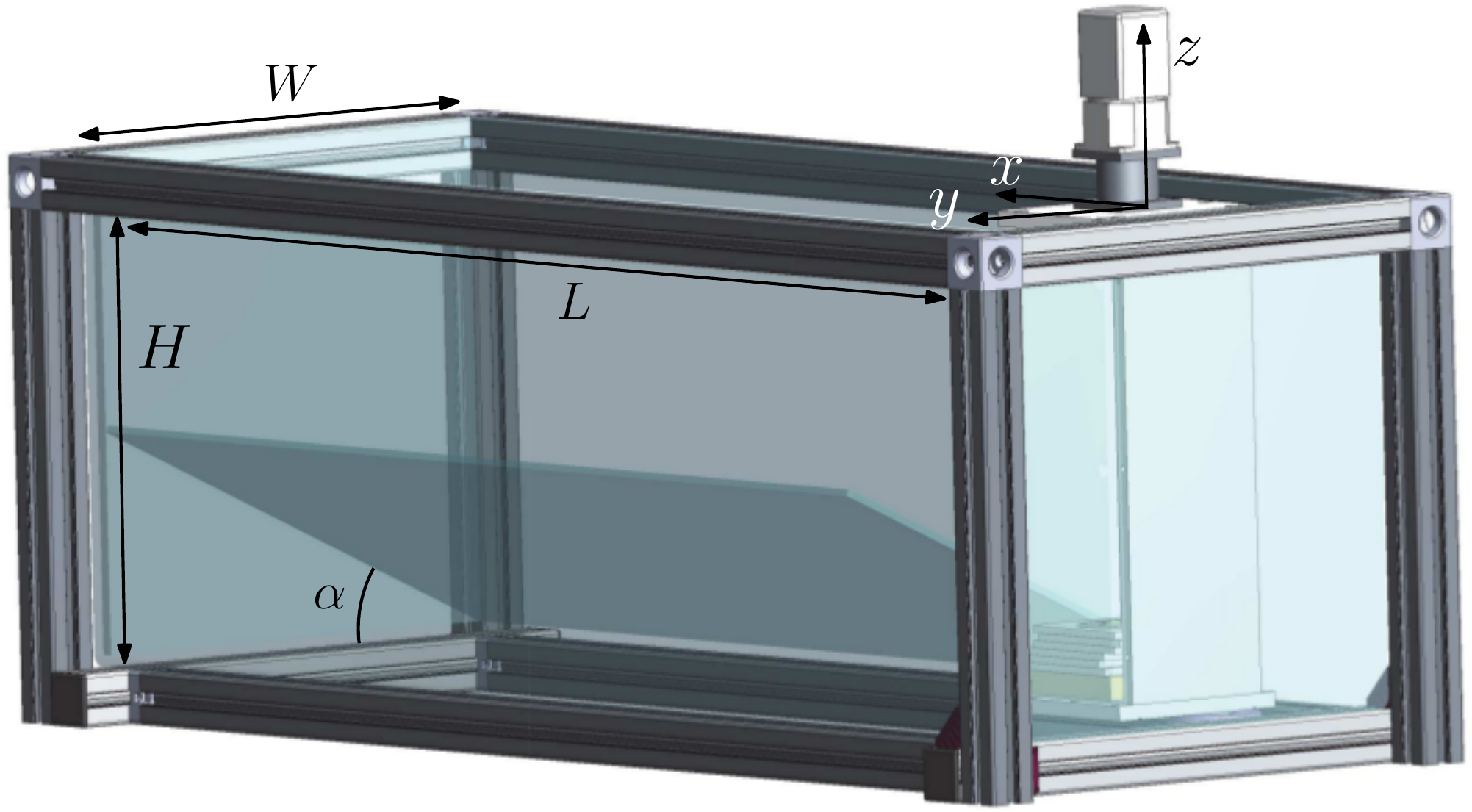}
\end{minipage}
\begin{minipage}{0.49\linewidth}
\centering
\includegraphics[width=0.5\linewidth]{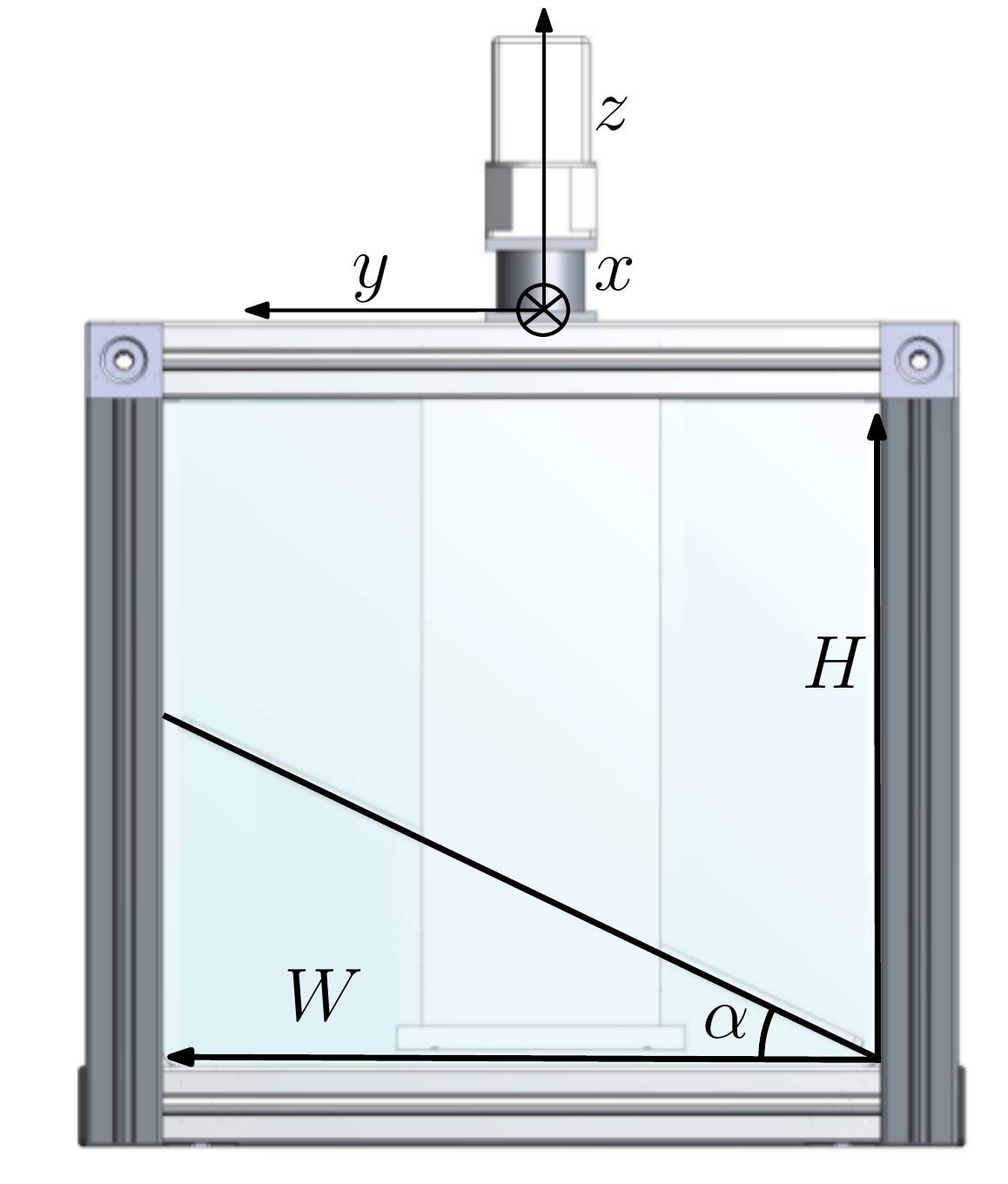}
\end{minipage}
\caption{The canal-like geometry: Perspective (left) and side (right) views of the experimental set-up. The slope and the wave generator are visible. The parameters used for the experiment are $H = 360$ mm, $L=1000$ mm, $W=410$ mm and $\alpha = 23\degree$. 
}
\label{exp2}
\end{figure}

\begin{figure}
\centering
\includegraphics[scale=1]{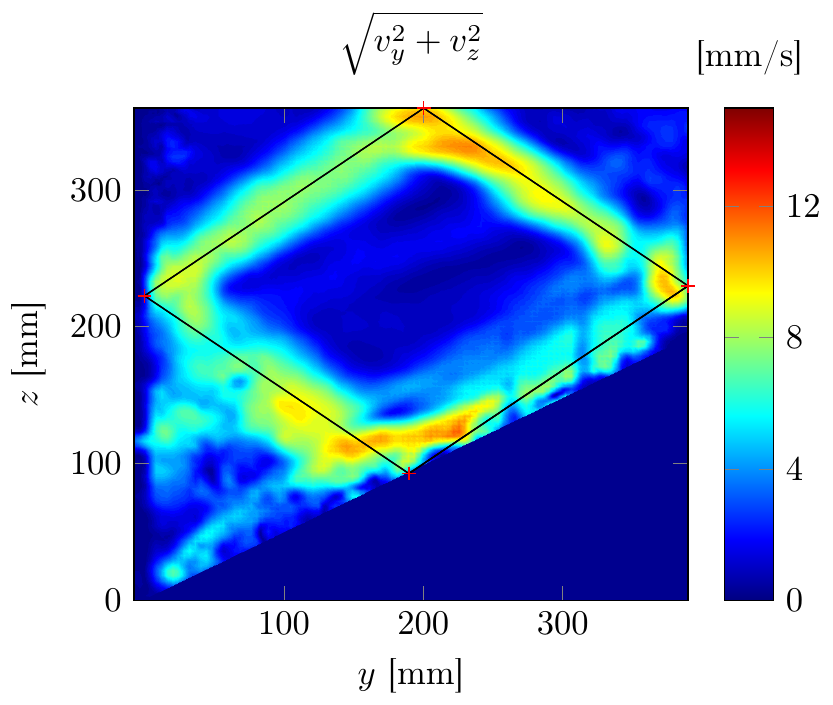}
\caption{Filtered velocity field amplitude $({v_y^2+v_z^2})^{1/2} $ at the slice $x$ = 700 mm. The ray tracing prediction is superimposed.}
\label{3Dattract_exp_th_2}
\end{figure}

\begin{figure}
\centering
\includegraphics[scale=1]{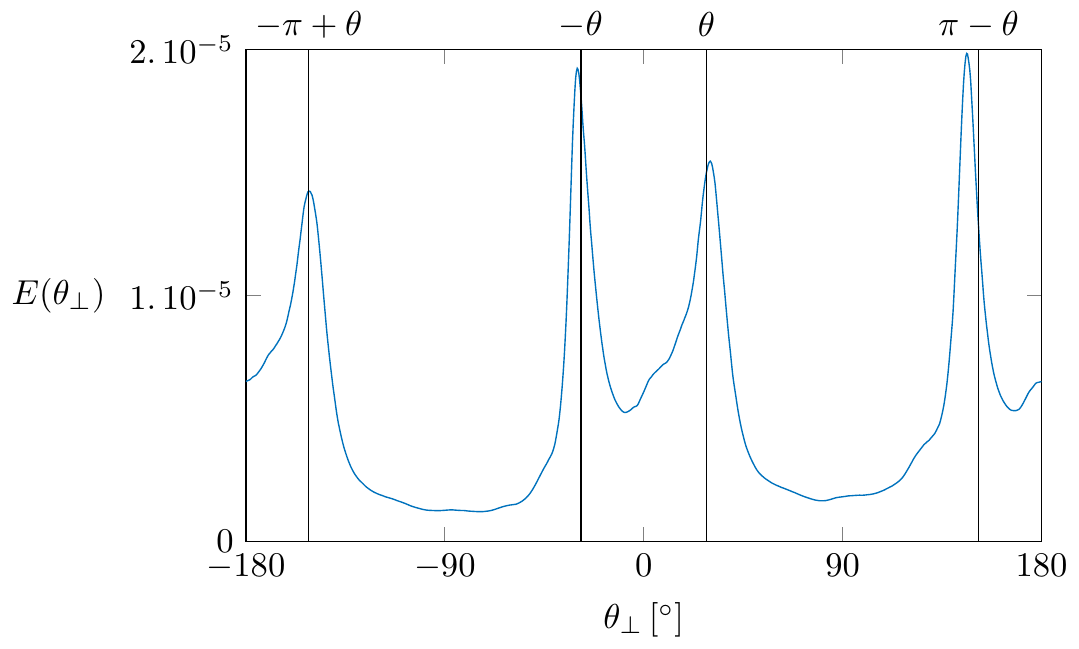}
\caption{Angular repartition of the energy density $E(\theta_\perp)$ for the filtered velocity field of the figure~\ref{3Dattract_exp_th_2}. The four black vertical lines represent the four $\theta$ values given by the dispersion relation determined by the forcing frequency.}
\label{spectre_ref}
\end{figure}

\begin{figure}
\centering
\includegraphics[scale=1]{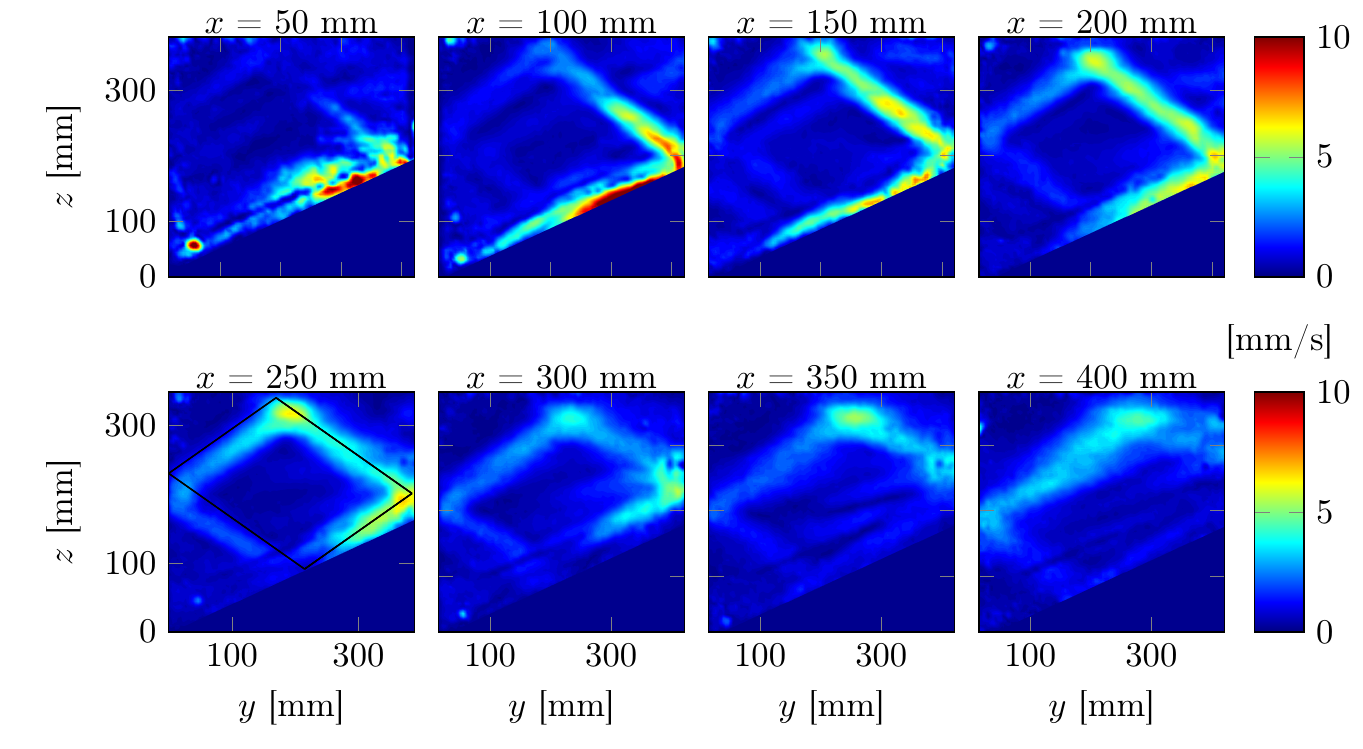}
  \caption{Filtered velocity amplitude $({v_y^2+v_z^2})^{1/2}$ for different slices of the tank.}
    \label{3D_tranches}
\end{figure}

For this geometry, two possible types of forcing were used :
\begin{itemize}
\item A mode~1~profile $\zeta(z,t) = a \sin(\omega t)\cos(\pi z/ H)${, limited to the dashed white region in Fig.~\ref{map_foc}} as previously, where $H$ is the total height of the fluid, $a$ the amplitude of motion of the cams, and $\omega$ the frequency of the forcing. This modal type of  forcing can be interpreted as sending of internal wave beams in both upward as well as downward direction. {Corresponding results are shown in Figs.~\ref{3Dattract_exp_th_2} and \ref{spectre_ref}.}

\item A plane-wave profile $\zeta(z,t) = a \sin(\omega t + 2\pi z/ \lambda)${, limited to the black rectangular region in Fig.~\ref{map_foc}} in which 
 the wavelength $\lambda$ and the numbers of wavelengths can be chosen. 
 We choose to force the fluid with only one wavelength. The upward phase propagation, adopted here, implies an inclined, downward-directed  internal wave beam is generated. {Corresponding results are shown in Fig.~\ref{3D_tranches}.}
\end{itemize}
    
For this experiment, a laser sheet is created above the tank that illuminates a transverse $yz$-section of the tank. The laser sheet can be translated to obtain different vertical slices of the tank. As before, we use the PIV method to obtain the velocity field. As previously, a Hilbert filtering is 
eventually used to filter the velocity field and to get rid of the noise.

In order to quantify the presence of the attractors and then conclude on the focusing effect predicted, we first force the fluid with a mode 1, the generator being put at the middle of the tank width. As we can see in figure~\ref{map_foc}, with this forcing {applied to the region inside the dashed white rectangle}, a large band of 150 mm$\times$300 mm in ($y_0$,$z_0$) is excited:  One can thus expect that such a forcing will create attractors approximately everywhere in the tank.
Experimentally, we do find attractor-like velocity fields over the whole length of the tank. 
Figure~\ref{3Dattract_exp_th_2} presents an example of the velocity field obtained for a slice where the amplitude is large: One clearly sees 
an attractor. The ray tracing prediction, calculated for the same parameters, has been superimposed on the experimental image, emphasizing the very good agreement.

To be quantitative and exhibit an experimental proof of the focusing of the energy in a transverse plane, we look at the wave propagation angles precisely in this transverse plane. Indeed,  PIV in a transverse laser sheet only measures
 the projection of the real velocity field in the laser sheet since the velocity perpendicular to the sheet (here, $v_x$, the velocity along $x$) is not observed. If $v_x$ 
 is small compared to the velocities in the other directions, then the propagation angle should follow the dispersion relation of internal waves 
$\omega = \pm N \sin \theta$.

In order to obtain the angles of propagation of the projected velocity field, we proceed in several steps that were already proposed in \cite{BrouzetEPL2016}. First, we filter the velocity field in $\omega$ using the Hilbert transform. Second, we perform the Fourier transform on $v_y$ and $v_z$, obtaining then the energy as a function of $k_y$ and $k_z$, two components of the wave vectors. We then interpolate $E(k_y,k_z)$
as a function of the perpendicular wavenumber $k_\perp=(k_y^2+k_z^2)^{1/2}$ and of the projected angle of propagation
$\theta_{\perp}={\mbox{arcsin} (k_z/k_\perp)}$ to obtain $E(k_\perp,\theta_{\perp})$. 
Integrating finally over all values of $k_\perp$, we get the energy density $E(\theta_\perp)$.

When applying this treatment to the velocity field shown in figure~\ref{3Dattract_exp_th_2}, we obtain the $E(\theta)$-repartition shown in figure~\ref{spectre_ref}.
The four black vertical lines represent the four $\theta$-values compatible with the dispersion relation  considering the experimental forcing frequency~$\omega$. One can see that the four peaks, representing the four branches of the attractor, fall on the four peaks theoretically expected.
So, as predicted by ray tracing, the focusing effect is taking place here.

To highlight the effect of energy trapping, due to this focusing enhanced in the ray tracing 
calculations, we now rely on a forcing with a plane wave. The goal is to create an attractor only in a given region of the canal to see, first propagating energy near the generator, then 
 attractors in the middle of the tank, and finally very low wave propagation beyond, due to the trapping of the waves. We choose a  rather large wavelength to limit the damping, but only one wavelength, 
to create attractors in a limited region only. The region of the forcing is shown in figure~\ref{map_foc} by the {region inside the black solid} contour.

Figure~\ref{3D_tranches} shows nicely the successful experiment. At the beginning ($x =$ 50 to 100 mm), one can see a shapeless velocity field; further downstream ($x=$ 150 to 300 mm),  the attractor is created and
especially visible in the two most energetic branches; beyond the focusing zone, the energy is very low.  Although the ray tracing predicted the attractor region around $x=$400 mm,
it appears to be located closer to $x=250$ mm. The difference can be explained by the strong sensitivity of figure~\ref{map_foc} on the parameters of the experiment.

Similarly to what has been performed for the reference case presented in figure~\ref{spectre_ref}, to get a more quantitative picture and show the energy focusing in a small region of the canal, it is useful to plot the position of the peaks of $E(\theta_\perp)$ obtained for different slices.
In this case, the procedure needs some care. For the first slices, waves propagate mainly through the observation planes. The {velocity field produced by} PIV is therefore not representative of the real 3D velocity field, hence the peaks are not clearly defined. One first step is to filter the velocity field in the $k_\perp$ space. Using again the Hilbert method, one can filter the velocity field for 
($k_y ,k_z > 0$), ($k_z < 0<k_y$), ($k_y < 0<k_z $), ($k_y,k_z < 0$), giving separately the four branches of the attractor. This procedure has proven its utility for the four branches of the St Andrew's cross~\citep{MGD2008} or for 2D attractors~\citep{BrouzetPRF17}. For each branch, one can moreover plot the angular repartition of the energy.

\begin{figure}
\includegraphics[scale=1]{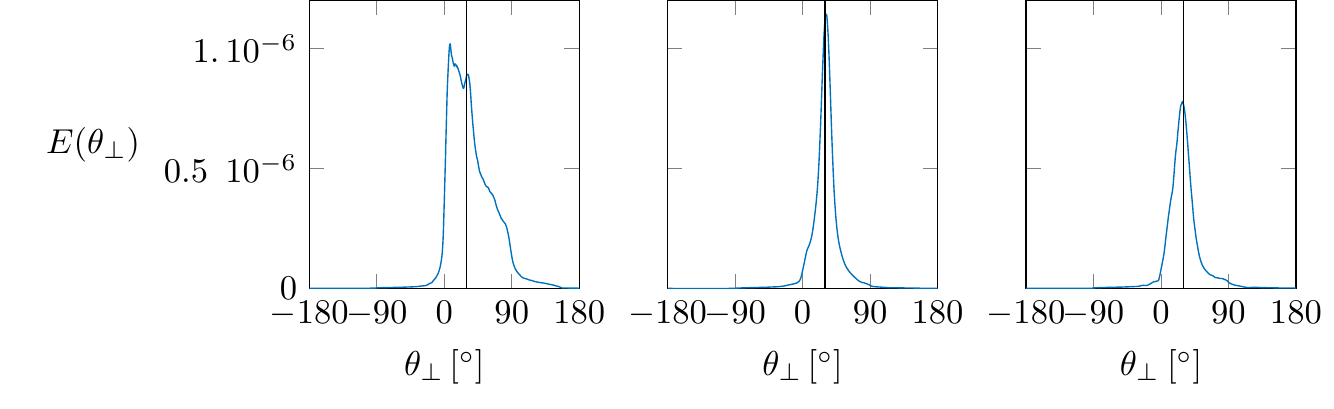}
\caption{Angular repartition of energy for the branch 3 corresponding to ($k_y > 0$, $k_z> 0$) 
for three slices. $x =$ 50 mm (Left); $x =$ 250 mm (Middle);  $x =$ 450 mm (Right).}
\label{3_branche3}
\end{figure}

\begin{figure}
\centering
\includegraphics[scale=1]{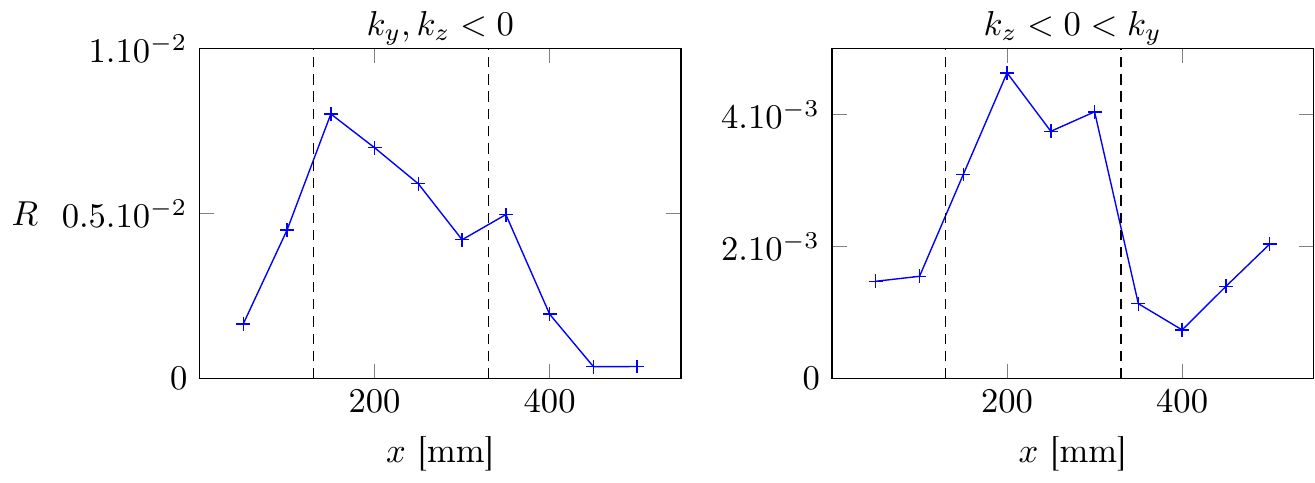}
\caption{Ratio $R$ of the peak amplitude divided by the total energy of a slice. The two graphs correspond to the two most energetic branches of the attractor, branch 1 (right) and branch 2 (left). The dashed lines represent the zone where $R > \max(R)/2$ for the right graph.}
\label{2b_ratio_peak_int}
\end{figure}

We thus study the positions of the peaks  and their shape along the longitudinal $x$-direction. For instance,  figure~\ref{3_branche3} presents the angular repartition of the energy for $k_y > 0$ and $k_z > 0$, i.e. in branch 3, counting counterclockwise from the attractor's slope-reflection, three different slices corresponding to three different $x$-positions. At the beginning 
and therefore close to the injection ($x$ = 50 mm, left graph), the energy is not focused on a single peak, and not even centered on the expected angle, represented by the solid vertical line: The wave is still propagating transversally to the observation slice. Further downstream
 in the canal ($x$ = 250 mm, center graph), the energy is almost centered along one single peak and its position corresponds to the angle given by the dispersion relation:  The focusing occurred and the waves are almost trapped in the observation slice. Even further downstream  ($x$ = 450 mm, right graph), the waves are still almost in the transverse plane, but the energy has significantly decreased, as the trapping occurred upstream.
Of course damping also plays a role to explain this decrease. That is why, to emphasize the focusing, one may
track the position of each of the four peaks, for all slices. However, as emphasized by the previous example, the peak position is not always the relevant quantity. 
The ratio of the peak amplitude to the total transversal slice energy
\begin{equation}
R =\dfrac{\max{(E(\theta_\perp))}}{\int_{\footnotesize{slice}}(v_y^2+v_z^2) \mbox{d}y \mbox{d}z}\label{DefinitionRatio}
\end{equation}
is a more appropriate indicator.
If this quantity is relatively large, it indicates that the energy is well focused around the main peak and therefore looking at the peak position makes sense. On the contrary if this quantity is low, one cannot clearly define a single direction of propagation in this region of the $k$-space and at this slice. The ratio~(\ref{DefinitionRatio})  can be calculated for all four branches of the attractor, but we chose to focus only on the two most energetic ones, obtained by filtering $k_\perp$ for branch 2 ($k_y ,k_z < 0$), 
 and for branch 1 ($k_z < 0<k_y $).  Figure~\ref{2b_ratio_peak_int} presents the value of $R$ for these two branches, for all slices.

Taking the most energetic branch (right graph of figure~\ref{2b_ratio_peak_int}), one can define a zone of confidence by the region for which $R > \max(R)/2$. This region, emphasized with the vertical dashed lines in figure~\ref{2b_ratio_peak_int} is approximately located between $x = 130$ mm and $x = 320$ mm.

\begin{figure}
\centering
\includegraphics[scale=1]{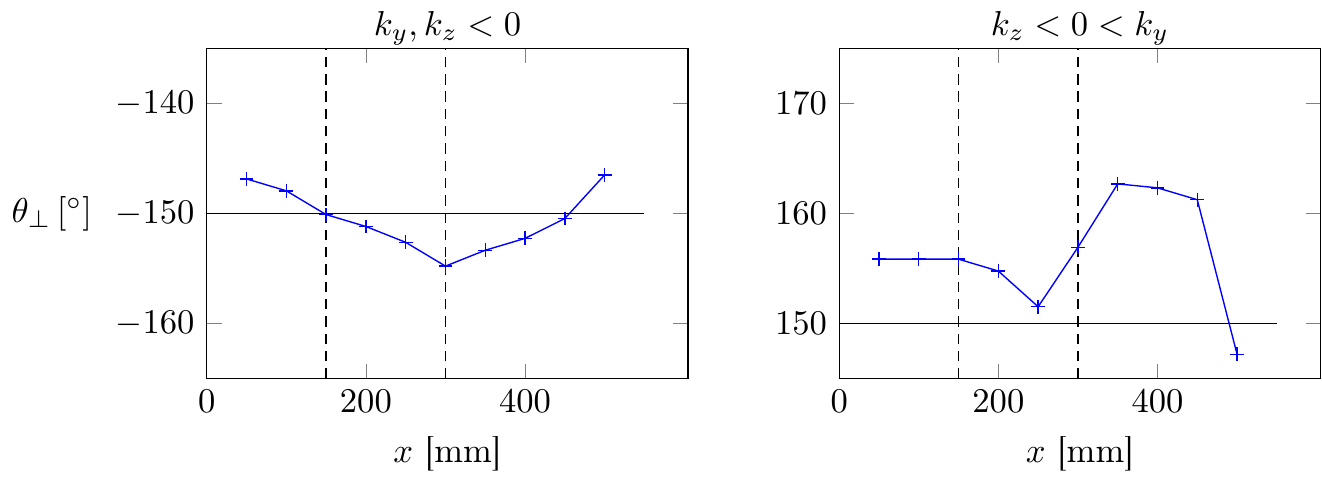}
\caption{Peak position for all slices, for the two most energetic branches. The vertical dashed lines delimit
 the region in which the energy is significantly focused on a single 
$\theta$-value, so where looking at the peak position makes sense. The horizontal black line shows the expected value of the propagation angle.}
\label{2b_theta_tranches}
\end{figure}

This quantity is not a proof of the focusing in itself but 
gives a good indication at which $x$-position along the canal
it is relevant to study the peaks position in the $(E,\theta)$ diagram. For these two branches, one can rely on the $E(\theta_\perp)$ peak position values if the slice is taken 
between the dashed lines. In this area, 
one can expect a good estimate of the transverse angle of propagation $\theta_\perp$. 
 In figure~\ref{2b_theta_tranches}, the peak position is plotted for all the slices, for the two branches. 
 We add moreover in each plot a horizontal line representing the expected propagation angle
 and two vertical dotted lines to show the confidence zone. For $x$-values in this zone, one  can see that the propagation angles correspond to the expected ones, confirming that, in these slices, the propagation is almost totally transverse:  trapping 
 occur{r}ed because of the focusing effect.

The ratio $R$ combined with the peak position for many $yz$-slices allow us to conclude that there is indeed a region approximately between $x = 150$ mm and $x = 300$ mm where  propagation of most of the energy is totally transverse. As  predicted by ray tracing, 
the refractive focusing mechanism is now experimentally confirmed.

\section{Conclusion}
\label{Conclusion}

We studied experimentally two simple 3D geometries. The first one is slightly different from a classical 2D study case for (1,1)-attractors. The difference lays in the forcing, which is  non-uniform, since we are forcing only over a 150 mm  wide lateral interval  in a 800 mm wide tank. The ray tracing in this geometry shows that if the rays are sent with a small angle with respect to the
{bottom-normal direction}, they are spreading everywhere throughout the tank, before ending up in a plane { oriented normal to the sloping bottom}, always forming attractors. This prediction was then confirmed experimentally. We showed first that the attractor velocity field is {nearly} identical everywhere in the tank {even though} forcing only occurs in the middle of the tank. In a subsequent experiment, we showed that attractors are created approximately 
everywhere almost simultaneously. Moreover, the energy was roughly uniformly spread throughout the tank, giving two strong arguments in favour of the proposed mechanism of generation { by refractive trapping}. 
{Complementary numerical experiments confirmed also the results, {revealed a phase shifting in the transversal direction} and provided in addition a very interesting confirmation of the role of boundary conditions. In particular, they emphasize
 the stronger {localized} response {near lateral walls} when using a no-slip instead of stress-free boundary condition. }

In the second experiment, the geometry of the tank is 
closer to  a canal having a sloping bottom. The forcing is applied in the along-slope, longitudinal  direction of the canal. The ray tracing and the experimental data showed unambiguously that the propagation ends up transversally to the canal. Experimentally, we proved the refractive focusing by looking at the angular repartition of the energy. We showed, indeed that, once focused, waves are propagating at the angles expected by the dispersion relation. We also showed that this focusing can significantly reduce the energy at the end of the canal, since waves are rapidly trapped in transverse planes. This may provide a tentative explanation for \textit{in situ} measurements of the Laurentian Channel, which show measurements of unexpectedly low internal wave energy far away from the channel head, unexplained yet~\citep{SaintLuarentAttenuation}.  Subsequent observations,  along one side of this Channel, do show evidence of internal tides. These appear to be forced at the nearby located transverse sill at the head of the Laurentian Channel, and while initially expected to propagate down-channel, they appear to be responsible for observed transverse internal tidal motions~\citep{Cyr2015}, such as expected from the refractive trapping mechanism.

This work shows the importance of the refractive focusing effect when considering 3D reflection of  internal gravity waves.
In both cases studied in this paper, the structure eventually created in the transverse planes is a 2D-like attractor. 
The parameter space diagrams suggest that attractors are indeed generic for these geometries. 
It is likely that there is a direct link between the focusing and the existence of attractors. 
Work along this line is in progress.

\begin{acknowledgments}
{\bf Acknowledgments}
This work was supported by the LABEX iMUST (ANR-10-LABX-0064) of Universit\'e de Lyon, within 
the program ``Investissements d'Avenir'' (ANR-11-IDEX-0007), operated by the French National 
Research Agency (ANR). This work has been supported by the ANR through grant ANR-17-CE30-0003 (DisET).
This work has achieved thanks to the resources of PSMN from ENS de Lyon. \end{acknowledgments}

\end{document}